\documentstyle[12pt]{article}
\title{IDEAS IN NONPERTURBATIVE QCD}
\author{}
\date{}
\newcommand{\be}{\begin{equation}}
\newcommand{\ee}{\end{equation}}
\begin{document}
\maketitle
\begin{center}

\vspace{-2cm}

{\it  Invited Talk\\
 at the International conference Hadron--93,\\
 21--25  June Como (Italy)}\\
\vspace{1cm}

{\bf Yu.A.Simonov\\
Institute of Theoretical and Experimental Physics\\ 117259, Moscow,
B.Cheremushkinskaya 25, Russia}
\end{center}

\begin{abstract}
The structure of the hadron spectrum is discussed in connection with the
main phenomena of nonperturbative QCD: confinement and chiral symmetry
breaking (CSB). For the higher part of the spectrum ($M \ge 2 GeV$) spin and
chiral effects are unimportant; spectrum of $q\bar{q}$ system is described
by an effective Hamiltonian deduced from QCD.  The Hamiltonian reduces  to
relativistic quark potential model or to the open string model in two
opposite limits. Hybrids are shown to appear naturally in theory and enter
the multiplets which are compared to experiment and bosonic string theory.
The phenomenon of conspiracy of the spectrum of radial excited states
producing operator product expansion is discussed.

The lower part of the spectrum $(M< 2 GeV)$ is influenced by spin and chiral
effects. CSB and the chiral quark mass are deduced for the vacuum containing
instantons and confining background.

Main points are summarized in conclusion.

\end{abstract}
\section{Introduction}

The main nonperturbative properties  of the QCD vacuum --
confinement and CSB -- influence  hadron spectrum in a different way for
higher $(M> 2 Gev)$ and lower $(M< 2 Gev)$ part.

 The former is shaped by confinement and mesons are mostly $q\bar{q}$
 connected by a string. The lower part feels strongly CSB especially in the
 PS channel.
 After a short discussion of the QCD vacuum in section 2 we devote next 4
 sections to derivation of $q\bar{q}$-string Hamiltonian from first
 principles and compare resulting multiplets to experiment. The
 last sections are devoted to chiral effects interconnected with
 confinement.  A summary of results is given in conclusion.

\section{Nonperturbative properties of the QCD}

 The QCD vacuum is known to be occupied by the
nonperturbative configurations, which lead to the scale
anomaly and produce the nonperturbative shift of the
vacuum energy density  $\varepsilon$ [1]
\be
\varepsilon = \frac{\beta(\alpha_s)}{16 \alpha_s} <F^a_{\mu\nu}
F^a_{\mu\nu}>\cong - \frac{11}{3}N_c \frac{\alpha_s}{32\pi}<F^2>
\ee

Note that  asymptotic freedom which ensures the negative sign of
$\beta(\alpha_s)$, makes the nonperturbative QCD vacuum advantageous as
compared to the empty (perturbative) one.

 The nonperturbative QCD vacuum can be characterized by  vacuum
field correlators\\ $<F(1)\Phi(1,2) F(2)...F(n)\Phi(n,1)>$
where $\Phi(x,y) $ are parallel transporters
\be
\Phi(x,y) = P ~exp~ig \int^x_y A_{\mu}(z) dz_{\mu}
\ee
The dynamics of confinement  enters
through the area law of the Wilson loop
\be
<W(c)> \equiv \frac{1}{N_c} <tr\Phi(x,x)>= exp (-\sigma S_{min})
\ee
where the string tension $\sigma$ is computed through the vacuum correlators
[2]
\be
\sigma =\frac{1}{24} \int d^2x g^2<F_{\mu\nu}(x)\Phi(x,0)
F_{\mu\nu}(0)\Phi(0,x)>+...
\ee
 and dots refer to higher order
correlators.

Another important characteristics of the nonperturbative vacuum
is the chiral condensate [3]
\be
<\bar{q}q> \cong - (250 MeV)^3
\ee
and the topological susceptibility [4]
\be
\int d^4x<Q(x)Q(0)> = (180 MeV)^4
\ee
which suggests that the topological charge density is of the order
 of one unit per $1 fm^4$. We shall see that the  latter  quantity is
connected to chiral condensate (5).

\section{The $q\bar{q}$ Green's function}

 The Green's function of the $q\bar{q}$ system can be written using the
Feynman-Schwinger representation [5] as a double path integral
over paths of a quark $Dz$ and antiquark $D\bar{z}$ with the proper-time
integrations $ds~d\bar{s}$
\be G(x\bar{x},y\bar{y}) = \int ds \int d\bar{s}
Dz D\bar{z} e^{-K-\bar{K}} <W(C)>
\ee
where we have omitted spin degrees of
 freedom having in mind to concentrate on higher levels, where spin
interactions are unimportant. We also neglected the quark determinant (sea
quark loops) in the large $N_c$ limit (quenched approximation).

Here kinetic energy terms are defined as
\be
K=\frac{1}{4} \int^s_0 \dot{z}^2_{\mu}(\lambda) d\lambda,
\bar{K}=\frac{1}{4} \int^{\bar{s}}_0 \dot{\bar{z}}^2_{\mu}(\lambda) d\lambda.
\ee
All interaction between $q$ and $\bar{q}$ is contained in the
Wilson loop in (7). To understand better the origin of confinement - the
area law (3) - one may apply to $W(C)$ the nonabelian Stokes theorem and
use the cluster expansion, which yields [2]
\begin{eqnarray}
<W(C)>= exp\{-\frac{g^2}{2!} \int
d\sigma_{\mu\nu}(u)d\sigma_{\rho\lambda}(u') <F_{\mu\nu}(u,z_0)
F_{\rho\lambda}(u'z_0)>+ \\ \nonumber +\frac{g^4}{4!}\int
d\sigma(1)d\sigma(2)d\sigma(3)d\sigma(4) \ll F(1,z_0) F(2,z_0)
F(3,z_0)F(4,z_0)\gg +...\}
\end{eqnarray}
where $F_{\mu\nu}(u,z_0)=\Phi(z_0,u)F_{\mu\nu}(u)\Phi(u,z_0)$ and $z_0$ is an
orbitrary
point, on which the whole sum (9) is independent.
It is convenient to choose it in the plane  of the contour C. It was shown in
[2]
that each term of the cluster expansion in (9) provides the area law (4)
when the area $S$ is much larger than the correlation length $T_g$ of
field correlators $<FF>,<FFFF>$ etc. Recently the lowest order correlator
$<FF>$ was measured on the lattice [6]. It consists of two independent
Lorentz structure functions $D$ and $D_1$, \begin{eqnarray} g^2<F_{\mu\nu}
(z,0)F_{\rho\lambda}(0,0)>= (\delta_{\mu \rho}\delta_{\nu\lambda}-
\delta_{\mu\lambda}\delta_{\nu\rho})D(z)+\\
\nonumber
+\frac{1}{2}[\partial_{\mu} z_{\rho} \delta_{\nu\lambda}-
\partial_{\mu} z_{\lambda} \delta_{\nu\rho}+
\mu\nu \leftrightarrow \rho\lambda]
D_1(z),
\end{eqnarray}
which both decrease exponentially [6]
\be
D_1(z)\sim \frac{1}{3} D(z)\approx const ~ exp (-|z|/T_g),~~ T_g \approx 0.2 fm
\ee
Therefore one may use the area law (3) to calculate spectrum of $q\bar{q}$,
when the size of the $q\bar{q}$ system $R$ is much larger than $T_g$,
\be
R\gg T_g
\ee
Condition (12) is fulfilled for most existing hadronic systems, except
for the ground state of bottomonium, where one must exploit for $<W(C)>$ the
more detailed form (9).  On the other hand, the condition (12) puts some
limits on the $q\bar{q}$ system considered as a string, as we shall see
below.

\section{The $q\bar{q}$ -string Hamiltonian}

 Our aim now is to calculate the spectrum of the $q\bar{q}$ system [7],
described
by (7). To this end we rewrite identically the kinetic terms (8)
\be
K+\bar{K}= \int^{T}_0 \frac{d\tau}{2}[\frac{m^2_1}{\mu_1(\tau)}+
\mu_1(\tau)(1+\dot{z}^2_i(\tau))+\frac{m^2_2}{\mu_2(\tau)}+\mu_2(\tau)(1+\dot{\bar{z}}
^2_i)]
\ee
introducing an important new quantity -- to be the dynamical $q$ and $\bar{q}$
masses:
\be
\mu_1(\tau)= \frac{dz_4(\lambda)}{d\lambda};~~
\mu_2(\tau)= \frac{d\bar{z}_4(\lambda)}{d\lambda},~
\tau\equiv z_4=\bar{z}_4
\ee
We choose $z_4=\bar{z}_4 \equiv \tau$ as an integration variable
in (13), and neglect the backtracking in time i.e. take $\mu_i(\tau) >0$.
Justification for it may be found on dynamical grounds - when coming back and
forth
in time, the quark is dragging with itself the heavy string; the action sharply
increases due to that, so such motion is dynamically suppressed.

An effective action for our system can be read off from Eq.(7) and (3).
\be
A=K+\bar{K}+ \sigma S_{min}
\ee
 where $S_{min}$ -- the minimal area inside the contour made of $q$
and $\bar{q}$ trajectories $z(\lambda)$ and $\bar{z}(\lambda)$ -- can be
constructed
 by connecting $z(\lambda)$ and $\bar{z}(\lambda)$ by straight lines:
\begin{eqnarray}
S_{min} = \sqrt{\dot{w}^2w^{'2}-(\dot{w}w')^2}~,~~ w_{\mu} = z_{\mu}\beta+
\bar{z}_{\mu}(1-\beta),\\
\nonumber
\dot{w}_{\mu} = \frac{\partial w_{\mu}}{\partial \tau}~,~~
w'_{\mu} = \frac{\partial w_{\mu}}{\partial \beta}=r_{\mu}
=z_{\mu}-\bar{z}_{\mu}.
\end{eqnarray}

The square-root form of $S_{min}$ can be eliminated in the standard way [7]
introducing the auxiliary functions $\nu(\tau,\beta)$ and
$\eta(\tau,\beta)$.  After integrating out the latter and the center-of-mass
coordinate $R_{\mu}$, one is left with the following effective action (we
consider equal mass case, $m_1=m_2=m$ and consequently, $\mu_1=\mu_2=\mu$)[7].
\begin{eqnarray}
A=\int^T_0
d\tau\{\frac{m^2}{\mu(\tau)}+\mu(\tau)+\frac{1}{2}[\frac{\mu(\tau)}{2}
\dot{\vec{r}}^2+\\
\nonumber
+\int^1_0 d\beta(\beta-\frac{1}{2})^2\nu(\tau,\beta)\frac{(\dot{\vec{r}}\times
\vec{r})^2}{\vec{r}^2}+
\sigma^2\bar{r}^2 \int^1_0\frac{ d\beta}{\nu} + \int^1_0
\nu d\beta] \}
\end{eqnarray}
The function $\nu$ introduced as an auxiliary function, actually has important
physical
meaning - it describes the energy density of the string.
We note first of  all that $\nu(\tau,\beta)$ and $\mu(\tau)$ have no
canonical momenta and should be found from the minimum of the effective action
(17).

Below we consider several limiting cases of (17) following
discussion given in [7].

i) nonrelativistic case,  $m\gg \sqrt{\sigma}$. One finds $\mu \sim m \gg \nu$,
in the leading order  $\mu(\tau)= m,~~\nu = \sigma |\vec{r}|$,
\be
A\cong \int^T_0 d\tau [2m_q+\frac{m_q}{4}\dot{\vec{r}}^2+ \sigma|\vec{r}|]
\ee

ii)relativistic case, $L=0$. The term $(\dot{\vec{r}}\times \vec{r})^2$
disappears in (17) and minimization of $\nu, \mu$ yields the Hamiltonian
\be H=2\sqrt{\bar{p}^2+m^2}+\sigma |\vec{r}|
\ee

This is exactly the Hamiltonian of the relativistic  potential model,
assumed in many  papers [8-10] and  studied both numerically
and  and quasiclassically in [9,10]. The spectrum is given  by a simple formula
\be
M^2(n_r,L)=4\pi\sigma (n_r+\frac{L}{2})+\Delta
\ee
where $\Delta$ has been computed numerically and quasiclassically in [9,10]
\be
\Delta(n_r,L)= 2\sigma(4-\pi - \gamma) L+ 2 m^2_q+4m^2_q ln \frac{M}{m_q}+
m_0^2
\ee
 $\gamma=0.12$ for $n_r$ large.

It is interesting that asymptotics at large $n_r$,$ M^2 \approx 4\pi\sigma
n_r$,
is twice that of bosonic string for large $L$, $M^2_L \approx 2\pi\sigma L$.
On the other hand at large $L$ the asymptotics of (20),
\be
M^2(L\gg 1)\approx 8\sigma L
\ee
differs from that of bosonic string.

Actually this happens because we have used in (22)
the Hamiltonian (19) in the region $L\gg 1$, where it is not applicable. To
find
out what regime takes place at large $L$, one must consider

iii) the limit of pure string dynamics, $L\gg 1$, $L\gg n_r$. In this case the
minimization in (17) yields $\nu\gg \mu$ and [7]
\be
M^2(L) = 2\pi\sigma L (1+(\frac{1.46(n_r+1)}{L})^{4/5} + ...
\ee
The extremal value of $\nu$ is the energy density of rotating string
\be
\nu \rightarrow \nu_L(\beta) = \frac{\rho_L}{\sqrt{1-v^2(\beta)}},
\ee
where the mass density of the string $\rho_L$ and the velocity $v(\beta)$
are given by \be \rho_L=
(\frac{8\sigma\sqrt{L(L+1)}}{\pi})^{1/2}~,~~v(\beta)=2(\beta-1/2)~, \ee
Thus one can see that $\mu(\tau)$ and $\nu(\tau,\beta)$ refer to the  energy
density of quarks and string respectively. At small $L$, we have
$<\mu>=<\nu>$ and $\nu$ describes the potential energy,
$\nu=\sigma|\vec{r}|$; $\mu$ is dynamical mass of quark. At large $L$ we
have $\nu \gg \mu$ and most energy is carried by string $(\nu)$ and not
by quarks $(\mu)$. This yields correct mass relation (23).

\section{Structure of the $q\bar{q}$ spectrum at $M\ge 2 GeV$}

One can write the general form of the spectrum, corresponding to the action
(17)
in the form (20) where $\Delta$ is given by (21) at moderate $L, L\le 2$
while at large $L$, the value of $\Delta$ is frozen, effectively one can
put $4-\pi-\gamma\rightarrow 0, L\rightarrow \infty$ in (21). This form
 of answer is also suggested  by numerical quantization of
the $q\bar{q}$ system in [11].

Since $\Delta$ is thus limited for large  $L$, the gross features of the
high excited spectrum are given by the
first term on the r.h.s. of (20), suggesting degeneration of states with
 the same $N\equiv n_r +L/2$.

Taking $\Delta(n_r,L)$ into account splits the masses within the
multiplet with a given value of $N$.

Qualitatively $\Delta$ grows with $L$ for $L$ not large; $L\ge 2$ and this
agrees
with experiment. E.g. for the doublet $N=1$
$\rho(1450)(n_r=1, L=0)$ and $\rho(1700)(n_r=0, L=2)$ we find from (21),
$\delta\equiv \Delta (1,0)-\Delta(0,2) \cong 0,58 GeV^2$ (we choose
$m_q=0.2 GeV$ and $\sigma=0.17 GeV^2$  , for discussion of $m_q$ see last
 section).

Experimentally $\delta_{exp} \equiv \Delta (1,0)-\Delta(0,2) \cong 0.79 GeV^2$.

Now for the triplet $N=2$
%\be
$$\rho_5(2350)(n_r=0, L=4), \rho_3(2250) (n_r=1, L=2), \rho_1(2150)(n_r=2,
L=0)$$
% \ee
 the theoretical difference $\delta$ between $\rho_5$ and $\rho_3$,
or $\rho_3$ and $\rho_1$ is again  $0.58 GeV^2$ (however it is actually
smaller for the first pair because for $L>2$ the value of $\Delta$ start to
saturate), while  experimentally $\delta_{53}=\delta_{31}= 0.46 GeV^2.$ We
list in Table 1 the theoretically computed masses using eqs. (20-21), and
compare these with experimental values. One can notice that agreement is
good, except for the lowest state -- $\rho$ and $\pi$ mesons which should
not be described by our formulas (20-21).

We conclude this section by several remarks:

1) lowest states need corrections from spin-spin, spin-orbit interactions and
gluon
exchanges, which are not taken into account in (20-21)

2) at large $L$, fixed $n_r, L\gg n_r$, the splitting $\Delta$ does not
depend on $L$ - this is the string regime,
[7] described by (23). E.g. from Table 1 one obtains that already $\rho_5$
state with $l=4$ is close to the string regime but correction in (23) is
already
 large,
\be
\Delta M^2 \approx 2\pi \sigma (\frac{1.46}{L})^{4/5} \cdot \sqrt{L(L+1)}
\approx 0.4 M^2 , (L=4)
\ee

3) at large $n_r$, fixed $L$, one has relativistic potential regime, $M^2=4\pi
\sigma n_r$.
It is
remarkable that the "radial trajectory" $\rho(2.11),\rho(1.45), \rho(0.77)$ has
a slope $4\pi\sigma$ different from lowest part of orbital Regge trajectory
$\rho(0.77), \rho_3(1.69), \rho_5(2.35)$ with the slope which is
close to $8\sigma$. This is in nice agreement with theoretical prediction (20).

4) There are many states especially in the $I=0$ channel, which do not fit into
the spectrum of multiplets $M^2=4\pi\sigma(n_r+L/2) +\Delta$. Some of
them --hybrids-- will be discussed in section 7. They fit into generalized
multiplets of the QCD string type.  Others  do not and they are probably not
 of the $q\bar{q}$ structure - they might be gluebals or multiquark states.
We obtain a fit of mesons in Table 1 taking  $m_q=0.2 GeV$. In section 9 we
justify  this choice showing that $m_q$ is actually not the current mass (as
$\bar{m}_q$ in (39)) but the chiral mass $M(0)$ (71).\\

 {\bf Table 1}\\

 \begin{tabular}{|l|l|l|l|}
 \hline
 &&&\\
 $L\backslash n_r$&0&1&2\\ \hline
 0&0.767&1.47&2.07\\
  &$\rho(0.77)$&$\rho(1.45);\pi(1300)$&$\rho(2.11);\pi(1.80?)$\\ \hline
 1&1.328&1.90&2.34\\
  &$a_2(1.32)$&&$f_2(2.34)$\\ \hline
2&1.715&2.19&2.58\\
 &$\rho_3(1.69),\pi_2(1670) \rho(1700)$&$\rho_3(2.25)$&\\ \hline
 3&2.029&2.44&\\
  &$a_4(2.04)a_3(2.05)$&&\\ \hline
 4&2.3~~~~~~~~~~~~~~~~~~~~2.31&&\\
  &$\rho_5(2.35)$&&\\ \hline
 5&2.544~~~~~~~~~~~~~~~~~~2.52&&\\
   &$a_6(2.45)$&&\\ \hline
   \end{tabular}          \\

   Masses of mesons computed from (20) (upper entry) $vs$ experimental
   values from Particle Data Booklet, June 1992 (lower entry). For $L=4.5$
   in the right upper corner are listed values computed in the corrected
   string regime, Eq. (23).

   Parameters used in Eq.(20-21) are:$m_q=0.2 GeV, (8\sigma)^{-1} = 0.85
   GeV^{-2},~ m_0^2$ was fitted to $\rho(0.77)$.

\section{OPE: condensates from hadronic spectra}

Consider as a first example the process $e^+e^- \rightarrow$ everything, its
crossection is
given by $Im\Pi_{\mu\nu}(q^2)  $, where $\Pi_{\mu\nu}
$ is
\be
\Pi^i_{\mu\nu}(x) = \frac{1}{2}<0|Tj^i_{\mu}(x)j^i_{\nu}(0)|0>
\ee
and $i$ denotes the sort of quark. One can write
\be
\Pi^i_{\mu\nu}(Q^2) = (Q_{\mu}Q_{\nu}-Q^2\delta_{\mu\nu})\Pi^i(Q^2),
\ee
\be
\Pi^i(Q^2) = \frac{1}{\pi}\int^{\infty}_{s_0}\frac{dsJm\Pi^i(s)}{s+Q^2},
\ee
Introducing the standard hadronic ratio $R_i(s)$, one has
\be
Im\Pi^i(s) = \frac{1}{12\pi ~e^2_i}R_i(s)
\ee
Consider now large $N_c,N_c\rightarrow \infty$. This is a realistic
limit, since it
provides a linearity of Regge trajectories
which is observed with a few percent accuracy. Also, all hadronic masses are
constant in this limit, while decay widths are $0(1/N_c)$ and are
indeed smaller than masses (at large masses $\Gamma/M \sim 10 \%)$.

It is important that $\Pi^i$ contains only poles when $N_c \rightarrow \infty$
\be
\Pi_i(Q^2) = \frac{1}{12\pi^2}\sum ^{\infty}_{n=0} \frac{c_n^i}{M^2_n+Q^2}
\ee
while
\be
R_i(s) = \sum^{\infty}_{n=0}c_n^i \delta(s-M^2_n)
\ee
As symptotically at large $s$ the quark-hadron duality tells us that the
averaged $R_i(s)$ is constant \be \int_{\Delta s}R_i(s) ds=e^2_iN_c\Delta s=
 \sum_{n<\Delta s}c_n^i = c^i_{\infty}\Delta n
\ee
Hence the quark-hadron duality (QHD) means that [9,12]
\be
 c^i_{\infty}= e^2_i N_c\frac{dM^2_n}{dn}
\ee
Let us check it now with $M^2_n$ and $c_n$ for our Hamiltonian (19).
The $c_n^i$ have been computed quasiclassically in [9]:
\be
\mbox{For} L=0, c^i_{\infty}= N_c e^2_i\frac{2}{3}m^2
,m^2=4\pi\sigma
\ee
\be
\mbox{For} L=2, \bar{ c}^i_{\infty}= N_c e^2_i\frac{1}{3}m^2
\ee

Since $M^2(n_r,L=0)=M^2(n_r-1,L=2)$, both states asymptotically degenerate and
one
has
\be
 c^i_{\infty}= N_c e^2_i(\frac{2}{3}m^2+\frac{1}{3}m^2)=N_ce^2_im^2,
\ee
while from (20) for large $n\equiv n_r$ one has
\be
\frac{dM^2_n}{dn}=m^2\equiv 4\pi\sigma
\ee
Hence the QHD (34) is asymptotically satisfied, as was recognized in [9].

Let   us now make a step forward: we can expand (31) at large $Q^2$ in powers
of $1/Q^2$
using for the l.h.s. the operator product expansion (OPE) (this idea without
reference to large $N_c$ and in a bit different setting has
been first used in [12].). Using OPE from [1] one gets
\begin{eqnarray}
\Pi^i_{\mu\mu}(Q^2)=
\frac{Q^2}{4\pi} \{ \frac{c^i_0}{M^2_0+Q^2}+\frac{c^i_1}{M^2_1+Q^2}+
\sum^{\infty}_{n=2}\frac{c^i_n}{M^2+Q^2}\}= \\
\nonumber
\frac{e^2_iN_c}{4\pi^2} \{-(1+\frac{\alpha_s}{\pi})ln \frac{Q^2}{\mu^2}+
\frac{24\pi^2\bar{m}^2_q}{Q^2}+\frac{8\pi^2\bar{m}_q<\bar{q}q>}{Q^4} +\\
\nonumber
+\frac{\pi^2}{3Q^4}\frac{\alpha_s}{\pi}<G^aG^a> - \frac{8\pi^3 \alpha_s}{Q^6}
[<j^a_{\mu 5}j^a_{\mu 5}>+\frac{2}{9} <j^a_{\mu} j^a_{\mu}>]
\end{eqnarray}

We have separated out the first two poles to approximate $c^i_n= c^i_{\infty}$
for
$n \ge 2$. The sum is  then
\be
\sum^{\infty}_{n=n_0}\frac{1}{M^2_n +Q^2}= -\frac{1}{m^2}\Psi
(\frac{Q^2+\Delta+n_0m^2}{m^2})+ divergent~~ const.
\ee
where $\Psi(z)=\frac{\Gamma'(z)}{\Gamma(z)}$ and has an asymptotic expansion
\be
\Psi (z)_{z\rightarrow \infty} =
ln z- \frac{1}{2z} - \sum^{\infty}_{k=1} \frac{B_{2k}}{2k z^{2k}}
\ee
with $B_n$ Bernulli numbers, $B_2=\frac{1}{6}$.

Using (41) in (39) one obtains equations, of which we quote only those
resulting from terms $ln Q^2 , 1/Q^2$ and $1/Q^4$
\be
c_{\infty}^i = N_c e^2_i m^2 (1+\frac{\alpha_s}{\pi})
\ee
\be
c_0^i+c^i_1  = c^i_{\infty}(\frac{\Delta+2m^2}{m^2}-\frac{1}{2})
\ee
\be
c_0^iM_0^2+c_1^iM_1^2 = - e^2_i \alpha_s\pi<G^aG^a>+\frac{c^i_{\infty}}{2m^2}
[(\Delta
+2m^2)(\Delta+M^2) +  B_2m^4]
\ee

One deduce from (39) and (42-44) that\\

1) logarithmic term in OPE is naturally emerging from the sum of high excited
states, also with correct coefficient if (42) is satisfied (this is QHD).\\
2) In the limit $\bar{m}_q\rightarrow 0$ ($\bar{m}_q$ -current mass) the OPE
has no
$1/Q^2$ term, while the sum (31) generally contains it. However, if
$\Delta = \frac{1}{2}m^2$, then $\Psi(\frac{Q^2+m^2+\Delta}{m^2})$ contains no
terms
of $1/Q^2$ in agreement with OPE. This case we shall call "ideal spectrum".\\
3) Assuming the states with $n=0,1$ nonasymptotic as in (39), one gets
from (44) that the lowest value of gluonic condensate is
$\frac{\alpha_s}{\pi}<G^aG^a> \approx 0.1 GeV^4$ i.e. around 8 times the
standard value of [1]. One should have in mind that the limit $N_c\rightarrow
\infty$
cuts off the quark loops and hence changes gluonic condensate,
which can be several times larger than the standard value.

The same type of estimates have been obtained in [12] both from heavy quarkonia
and
 light quark channels.\\
4) Expansion (41) is at best asymtotic, since $B_{2k}$ grows as $k!$ at large
$k$.
Hence the OPE is at best asymptotic expansion with factorially growing
coefficients
of $(1/Q^2)^n$. \\
5) In the leading order of $N_c\rightarrow \infty $ the relation (39) is exact.
It allows to relate the spectrum in each channel $J^{PC}$ to the microscopic
characteristics of the vacuum -- condensates.
Condensates are the same in each channel; therefore one has very rigid
conditions on masses and coefficients $c_n$ in each channel. It may lead
to an apparent paradox. E.g. in the $1^{++}$ shannel the OPE for
$\bar{m}_q=0$ looks the same as in the $1^{--}$ channel (up to terms
$1/Q^4$). However the spectrum at least for $M<2 GeV $ looks very different.
What is the resolution of this paradox is not yet clear.

\section{Hybrids}

To define hybrids one has to separate quantum gluon field $a_{\mu}$ from the
nonperturbative background $B_{\mu}$
\be
A_{\mu}=B_{\mu}+a_{\mu},
\ee
with the background gauge condition $D_{\mu}(B) a_{\mu} =0$. Then the hybrid
state
w.f. can be formed as
\be
\Psi(x,\bar{x},u) = \bar{\Psi}(\bar{x}) \Phi(\bar{x},u) \Gamma a_{\mu}(u)
\Phi(u,x) \Psi(x)
\ee
Thus hybrid is obtaind as a product of the quark  bilinear $\bar{\Psi}\Gamma
\Psi$
with quantum numbers $0^{++}(\Gamma=1), 1^{--}(\Gamma = \gamma_{\nu}) ,
  0^{-+}(\Gamma= \gamma_5)$ and gluon w.f. with $J^{PC}=1^{--}$. As a result
one
gets for the hybrid
$$J^{PC}(\Gamma a_{\mu})= 1^{--} (\Gamma a_{\mu}= a_{\mu}), 0^{++} (\Gamma
a_{\mu }=
\hat{a}), 1^{+-}(\Gamma a_{\mu} = \gamma_5 a_{\mu})$$
$$1^{++}(\Gamma a_{\mu}= \sigma_{\mu\nu} a_{\nu}),2^{++} (\Gamma a_{\mu}=
\gamma_{\mu}a_{\nu} +  \gamma_\nu a_{\mu}) $$

Those are the lowest states not containing orbital gluon excitations.
Introducing into $\Gamma$ the operator $D_{\mu}(B)$ one gets all
possible gluon excitations with additional quantum
numbers e.g. $1^{-+}(\Gamma a_{\mu} = \gamma_{\nu} D_{\nu} a_{\mu})$.
Lowest states $1^{+-}, 0^{++}, 1^{++}, 2^{++} $ are degenerate modulo spin-spin
interactions of quarks and gluons.

The hybrid Green's function can be written  in the  same way as for $q\bar{q}$
system (cf.(7))
\be
G(1;2) = \int \prod^3_{i=1} (ds_iDz_i e^{-K_{i}})<W(C)\Phi_{adj}(u,v)>
\ee
where $W(C)$ is the product of quark parallel transporters $\Phi$, while
$\Phi_{adj}$ arises from the gluon propagator.

In the large $N_c$ limit the gluon line $\Phi_{adj} $ becomes a
$q\bar{q}$ line,  and we have
\be
<W(c)\Phi_{adj}>_{N_c \rightarrow \infty} \rightarrow
<W(C_1)><W(C_2)>
\ee
Thus the gluon line becomes a border of two surfaces $S_1, S_2$
and lies entirely inside the film covering the total contour $C$.
{\underline This means that gluon in the hybrid describes (at least at $N_c
\rightarrow \infty$) vibration of the surface and in this way vibrational
degrees
of freedom of the string appear, which have been absent in the
ground state (where the minimal surface enters).} We shall come
back to this point in the next section.

It is easy to obtain  from (47) the Hamiltonian in the same way as it was
 done for the $q\bar{q}$ system.
For small orbital momenta one gets
\be
H=\sqrt{\bar{p}^2_1 + m^2_1}+\sqrt{\bar{p}^2_2 + m^2_2}+ |\bar{p}_3|
+\sigma |\bar{r}_1-\bar{r}_3|+
+\sigma |\bar{r}_2-\bar{r}_3|-C_0
\ee
Calculation witout one-gluon exchange (OGE) and $\sigma= 0.17 GeV^2$ yields
lowest mass
\be
M=2.5 GeV - C_0=0
\ee
and including $OGE$ and $C_0=0.4\div 0.2 GeV$ [5] one gets
\be
M=1.3-1.5 GeV
\ee
This agrees qualitatively with other calculations [13]. Orbital gluon
excitations cost $\Delta M\approx 1 GeV$ for $L=1$.

\section{QCD string, bosonic string and Veneziano spectrum.}

The spectrum of the open bosonic string is [14]
\be
M^2_n = 2\pi\sigma [-\alpha_0+ \sum^{\infty}_{k=1} kN_k]
\ee
where $k$ denotes the mode and $N_k$ - excitation number.
Theory is consistent when $\alpha_0=1, d=26$.

The lowest mode is $k=1$, which is rotation of a rigid stick with $L=1$.
Next is $k=2$ which may be rotation with $L=2$ and vibration--center of the
string moves with respect to ends. The lowest vibration mode has
$M^2_2=4\pi\sigma\equiv m^2$. The open bosonic string has no longitudinal
(radial) excitations.

An important characteristics of the spectrum is the multiplicity of the
state with  given $N$, which is equal to $exp(a\sqrt{N}), N\rightarrow
\infty, a=\frac {4\pi}{\sqrt{6}}$ [14].  This exponential growth is needed
to get the Veneziano formula for the amplitude. In other words the property
of duality of amplitudes $A(s,t,u)$ which is contained in the Veneziano
formula, needs the exponential growth of multiplicity.

It is clear that the spectrum (20) cannot ensure this exponential growth -- the
number of states with given $n_r,L$ grows only like a power, because number
of degrees of freedom is fixed. We shall see now that the
problem is solved by hybrids.

One can recognize in the hybrid Hamiltonian (49) two pieces of string connected
at the gluon position. If one does the same type of treatment as for the
$q\bar{q}$
state leading in that case to (17) and considers a generic string excitation
with
spectrum (23) for each peace of string, one has for the asymptotic hybrid
spectrum
\be
M^2(l_1,l_2)= 2\pi\sigma(|l_1|+|l_2|), \vec{L}
=\vec{l}_1+\vec{l}_2
\ee
In case of pure vibration $
\vec{l}_1+\vec{l}_2 =0, |\vec{l}_1|=|\vec{l}_2|=\nu$ and one has
\be
M(\nu)=4\pi\sigma\nu,~~\nu=0,1,2...
\ee
where $\nu$ refers to the vibration mode.

For a multihybrid with $n$ gluons sitting on the $q\bar{q}$ string dividing
it into $n+1$ cuts, the vibration obtains when internal cuts have  angular
 momentum $2l$ while first and last have $l$.  The total mass again in
the regime when $\nu_i\gg\mu_i$ (string regime [7]) is
\be
 M^2(n,l) =
4\pi\sigma nl
\ee
 Thus every quantum of vibration yields $4\pi\sigma = m^2$
to the squared mass, while every quantum of rotation is $2\pi\sigma$.

The hybrids contribute all necessary vibration modes and correspond to the
spectrum (52). Moreover, the multiplicity is now growing exponentially,
since the number of degrees of freedom contains an infinite number of gluons
on the string.

The hybrids enter the same QCD string multiplets, which we can now write as
\be
M^2(n_r,L\nu)=4\pi\sigma(n_r+\frac{L}{2}+\nu)+\Delta
\ee
where $\nu\ge $ number of gluons in the multihybrid.\\
For $\nu=1,~~M_1=\sqrt{(1.46)^2+\Delta}\approx 1.5 GeV$\\
For $\nu=2,~~M_2=\sqrt{(2.06)^2+\Delta}\approx 2.2 GeV$

Conclusions on high spectrum:\\
Spectrum consists of QCD-string multiplets (56), containing radial, orbital
and vibrational exitation. This spectrum contains that of the bosonic string
plus radial excitations specific for $QCD_2$.

However, there is a limitation -- the finite correlation length $T_g$ (see
eg. (11)) makes a natural cut-off at small distances -- there is no string
at distances $\Delta x < T_g$. Therefore effective number of gluons is less
than length of the string divided by $T_g$. Hence the effective number of
degrees of freedom is finite and the string theory is nonlocal.
This fact
may cure difficulties with string quantization for the real QCD string for
$d=4$.

 \section{Lowest states -- chiral effects.}

 In the formation of lowest states the broken chiral symmetry plays an
 important role. In this Section  we discuss chiral quark mass and chiral
 symmetry breaking (CSB) in connection with confinement.

 Several statements known in literature are in order.

 1) CSB is due to quasizero modes of quarks in the vacuum gluonic field
 [15]. If $u_n, \Lambda_n$ are to be found from
 \be
 i\hat{D}(A) u_n(x) = \Lambda_n u_n(x)
 \ee
 then the quark condensate is connected to the density $\nu(\Lambda)$ of
 quasizero modes [15]
 \be
 <\bar{\Psi}\Psi> = -m_q\int\frac{\nu(\Lambda)d\Lambda}{\Lambda^2
 +m^2_q}\rightarrow -\nu(0)\pi
 \ee

 2) zero modes $u_0(x)$ are provided by instantons [16]. Therefore instantons
 may be responsible for CSB [17]. This is the simplest possibility. Another
 is vacuum with magnetic monopoles which also provide zero models [18]. The
 quark zero mode on (anti) instanton is normalizable [16].
 \be
 u^{(\pm)}_0=\frac{\rho}{\pi(\rho^2+ x^2)^{3/2}}\frac{\hat{x}}{\sqrt{x^2}}
 (^{~1}_{-1})\varphi,~~\varphi_{am}=\frac{1}{\sqrt{2}}\varepsilon_{am}
 \ee
 where $\rho$ is the size of instanton; $a,m$-color and spin indices.

 3) To provide $\nu(0)\not= 0$ and consequently CSB it is
 necessary for instantons and antiinstantons in the vacuum to
 overlap their zero modes [19]. This mechanism was realized in [19]
 neglecting gauge invariance and confinement properties. Here $I$ wil quote
 results [20] taking these properties into account.

 Consider instantons in the confining background
 \be
 A_{\mu}= B_{\mu} + \sum^N_{i=1} A^i_{\mu}~,~~ N=N_++N_-
 \ee
 where $B_{\mu}$ ensures confinement (i.e. correlators $F_{\mu\nu}(B) $ in
 (4) yield nonzero string tension), while $A_{\mu}$ is the field of $i$-th
 (anti) instanton (instantons do not confine).
 Now let light quark move in the field (60) where instantons are at $x=R_i,
 i=1,...N$. The scattering amplitude of a quark on center $i$ (instanton or
 antiinstanton) is given by
 \be
 t_i = \frac{u_0(x-R_i) u_0^+(y-R_i)}{i\bar{m}_q}
 \ee
 where $u_0$ is given in (59), $\bar{m}_q$- the current quark mass.
 The total quark Green's function is given by the multiple scattering theory
 as [19]
 \be
 S(x,y) = S_0(x,y)+\sum_{i,j} u^i(x) (\frac{1}{\hat{T}-i\bar{m}_q})_{ij}
 u^{j+}(y)
  \ee
   $$u_i\equiv u_0(x-R_i)$$ where $T_{ij}$ is the overlap
 integral of zero modes \be T_{ij}= \int u_i^+(z) i\hat{D}(B) u_j(z) d^4z
 \ee

 Eq.(62) shows that multiple scattering can provide effective quark mass --
 chiral mass. In solids this effective mass is naturally produced by
 subsequent collisions. For  CSB this is not enough -- the
 quark should return to a given center any number of times.
 Only in this case occurs a gap equation yielding the chiral mass (similar
 conclusions are drawn in [19] without confining field $B_\mu$). In case of
 one  flavour, $N_f=1$, the  effective action for the  quark can be written
 in the gauge-invariant way [20]
 \be
 Z_{QCD}= const\int _{\mu}D_{\mu}(B)D{\Psi}D{\Psi^+}exp\int
 dxdy\Psi^+(x)[i\hat{D}\delta(x,y)+iM(x,y)]\Psi(y)
 \ee
 where the nonlocal mass operator is
 \begin{eqnarray}
 M(x,y)=\frac{\varepsilon N}{2N_c V}\int dR i\hat{D}u_+(x-R)\Phi(x,R,y)u^+_+
 (y-R)i\hat{D}+ \\
 \nonumber
 +(+\rightarrow -) \equiv M_+ + M_{-}
 \end{eqnarray}
 and $\Phi(x,R,y)\equiv\Phi(x,R)\Phi(R,y)$ is product of parallel
 transporters (3).

 Parameter $\varepsilon$ is to be defined from the "gap equation", which is
 gauge invariant and contains confinement.
 \be
 \frac{N}{2}= Tr(\frac{M_+ M_+}{-\hat{D}^2 + M_+ M_{-}})
 \ee

 When no confinement is taken into account, $B_{\mu} = 0$, one can introduce
 $M(p)$ instead of $M(x,y)$ and (66) becomes [19]
 \be
 \int\frac{d^4 p}{(2\pi)^4} \frac{M^2(p)}{p^2 +M^2(p)} \frac{4VN_c}{N} = 1
 \ee
 with
 \be
 M(p)=\frac{\varepsilon N}{2VN_c}p^2\varphi^2(p)
 \ee
 and $\varphi(p)$ -- Fourier transform of the spacial part of $u_0(x)$
 (59).  One can find $\frac{N}{V}=R^{-4}$ from gluonic condensate [1]
  \be
 \frac{\rho}{R}=\frac{1}{3} \;,\; \frac{N}{V}=\frac{<G_a G_a>}{32\pi^2}=1
 fm^{-4},
 ~~\mbox{and}~~ M(p=0)=345 MeV
 \ee

 $M(p)$ is fast decreasing for large $p$; $<\bar{\Psi}\Psi> =-(255 MeV)^3$
 [19].

 How confinement modifies this picture? First, the density of instantons
 $d(\rho)$ is suppressed at large $\rho$ due to the freezing of the coupling
 constant $\alpha_s(\rho)$ at large distances in the confining background
 [21]. Rough estimates yield the average instanton size $\rho\approx 0.2 fm$
 [22].

 Second, the density of instantons $N/V$ decreases since now only a part of
 gluonic condensate is due to instantons.

 Keeping $<\bar{\Psi}\Psi>$ at experimental value, $-(250 MeV)^3$, one gets
 roughly [22]
 \be
 R\approx 1.2 fm \;\; , \;\; \rho = 0.22 fm \;\; \mbox{and}~~~ M(0) = 0.2
 GeV .  \ee It is interesting that for such  small instantons there appears
 a situation with two scales [23]\\ chiral scale $R_{ch} \approx \rho
\approx 0.2 fm$  and confinement scale $R\ge \sigma^{-1/2} \sim 0.5
fm - 1 fm$.\\

In the limit $R_{c}\gg R_{ch}$ one obtains that the mass operator becomes
local, and the role of the chiral mass is played by $M(0)$, where $M(p)$ is
given in (68). Thus, confinement and chiral effects are separated:\\
1) chiral mass is created at small distances, $x\sim \rho\sim 0.2 fm$ due to
quark returns to instantons while passing the vacuum.\\
2) at large distances, $x>T_g, q$ and $\bar{q}$ form a string, which is
described by the Hamiltonian (19)
(for $L\le n_r$) or effective action (17), where now the role of mass in
$m$ is played by $M(0) = 0.2 GeV$.
The situation with chiral mass is the same for more flavours. E.g. for
two flavours one makes bosonization (introduces auxiliary scalar  and
pseudoscalar fiels $\sigma,\eta,\sigma_i,\pi_i$ to disentangle  4
fermion vertices [19]). Integrating out quark and boson fields one gets
the gauge invariant effective Lagrangian for pions in the background
field $B_{\mu}$
\be
W(\pi) = - Tr ln(i\hat{D}(B)+i\hat{M}V_5)
\ee
where $\hat{V}_5= exp i\pi_i\tau_i\gamma_5$, and $\hat{M}$ is given in (66).
In the limit $B_{\mu}\rightarrow 0$ one obtains the action studied in [19].

\section{Conclusions}

Nonperturbative QCD naturally explains confinement and CSB, and through this,
the structure of the meson spectrum both in high and low mass
region. It is remarkable that QCD forms multiplets, called the
QCD- srting multiplets which contain those of bosonic string
 plus radial excitations -- those of $QCD_2$.
 To make this contact with bosonic string one needs string vibrations -- and
 remarkably its QCD counterpart are hybrids.

 Their appears a remarkable conspiracy in structure of spectrum which yields
 through OPE microscopic characteristics of the QCD vacuum -- gluonic and
 quark condensates.

 Finally, CSB and confinement work together to provide two distinct scales,
 and the chiral mass of quarks appears naturally, which enters into
 Hamiltonian (19). Thus chiral mass is created "before the string appears
 between quarks".

 The author is grateful to A.M.Badalian, H.G.Dosh, A.Yu.Dubin
 B.V.Geshkenbein and A.B.Kaidalov for usefull discussions. The financial
 support of the Alexander von Humboldt Stiftung and Organizing Committee of
 Hadron-93 which made this talk possible is gratefully acknowledged.

 \end{document}